\documentclass[a4paper]{jpconf}

\usepackage{graphicx}

\begin{document}

\title{$\pi\pi$ and $\pi K$ scatterings in
three-flavour resummed chiral perturbation theory}

\author{S\'ebastien Descotes-Genon}

\address{Laboratoire de Physique Th\'eorique,\\ CNRS/Univ. Paris-Sud 11 (UMR 8627),
91405 Orsay Cedex, France}

\ead{descotes@th.u-psud.fr}

\begin{abstract}
The (light but not-so-light) strange quark may 
play a special role in the low-energy dynamics of QCD. The presence of
strange quark pairs in the sea may have a significant impact
of the pattern of chiral symmetry breaking : in particular 
large differences can occur between the chiral limits of 
two and three massless flavours (i.e., whether $m_s$ is kept at its physical
value or sent to zero). This may induce problems of convergence in 
three-flavour chiral expansions. To cope with such difficulties, we introduce a new framework, called Resummed Chiral Perturbation Theory. We exploit it to
analyse $\pi\pi$ and $\pi K$ scatterings and match them with
dispersive results in a frequentist framework. Constraints on three-flavour 
chiral order parameters are derived.

\end{abstract}

\section{Two chiral limits of interest}\label{sec:msdep}

Because of its intermediate mass,
the strange quark has a special status in low-energy QCD. 
It is light enough to allow for a combined expansion of
observables in powers of $m_u,m_d,m_s$ around the $N_f=3$
chiral limit (meaning 3 massless flavours): $m_u=m_d=m_s=0$.
But it is sufficiently heavy to induce significant 
changes from the $N_f=3$ limit to
the $N_f=2$ limit: $m_u=m_d=0$ and $m_s$ physical.
Each limit can engender its own version of
Chiral Perturbation Theory ($\chi$PT). 
In the $N_f=2$ limit, the pions are 
the only degrees of freedom, whereas $N_f=3$ $\chi$PT deals with
pions, kaons and $\eta$. This second version of $\chi$PT is richer, discusses 
more processes in a larger range of energy, but contains more 
unknown low-energy constants (LECs) and may have a slower convergence. 
The details of the connection between 
the two theories remain under debate.

Indeed, due to $\bar{s}s$ sea-pairs, order parameters
such as the quark condensate and the pseudoscalar decay constant,
$\Sigma(N_f)=-\lim_{N_f} \langle\bar{u}u\rangle$ and 
$F^2(N_f)=\lim_{N_f} F^2_\pi$,
can reach significantly different values in the two chiral limits 
($\lim_{N_f}$ denoting the chiral limit with $N_f$ massless 
flavours)~\cite{param}.
An illustation is provided by the quark condensate:
\begin{eqnarray}
\Sigma(2)&=&\Sigma(2;m_s)=\Sigma(2;0)
   + m_s \frac{\partial\Sigma(2;m_s)}{\partial m_s} + O(m_s^2)\\
&=&\Sigma(3)
   + m_s \lim_{m_u,m_d\to 0}
      i\int d^4x\ \langle 0|\bar{u}u(x)\, \bar{s}s(0) |0\rangle + O(m_s^2) \label{eq:sigma}
\end{eqnarray}
Here, $\bar{s}s$-pairs are involved through the two-point correlator 
$\langle (\bar{u}u) (\bar{s}s)\rangle$, which violates the Zweig rule in the vacuum (scalar) channel.
One expects~\cite{param} that this effect should
suppress order parameters when $m_s\to 0$: 
$\Sigma(2) \geq \Sigma(3)$ and 
$F^2(2) \geq  F^2(3)$. Since the quark condensate(s)
and the pseudoscalar decay constant(s) are the leading-order LECs
for the two versions of $\chi$PT,
a strong decrease from
$N_f=2$ to $3$ would have an impact on the 
structure of the two theories. 

\section{Two- and three-flavour chiral expansions}

A few years ago, the E865 collaboration 
provided new data on $K_{\ell 4}$ decays~\cite{E865}.
Building upon the dispersive analysis
of $\pi\pi$ scattering~\cite{royeq}, we extracted
the two-flavour order parameters~\cite{pipi}:
\begin{equation}
X(2)=(m_u+m_d)\Sigma(2)/(F_\pi^2M_\pi^2)=0.81\pm 0.07 \qquad
Z(2)=F^2(2)/F_\pi^2=0.89\pm 0.03
\end{equation}
A different analysis, with 
an additional theoretical input from the scalar radius of the pion, led to an even larger 
value for $X(2)$~\cite{CGL}. 
The situation is somewhat modified by new data from the NA48
collaboration~\cite{NA48}, which show some discrepancy with the E865 phase
shifts in the higher end of the allowed phase space. 
The role of isospin breaking corrections is under discussion currrently.
The preliminary
 values of the phase shifts~\cite{NA48} tend to increase the value
of the $I=J=0$ $\pi\pi$ scattering length, and to decrease the
value of the two-flavour quark condensate, pushing $X(2)$ down to 0.7.
In any case, one would expect values closer to 1, since 
$X(2)$ and $Z(2)$ monitor the convergence of $N_f=2$ chiral expansions
of $F_\pi^2 M_\pi^2$ and $F_\pi^2$ respectively. Such expansions
in powers of $m_u$ and $m_d$ \emph{only} should exhibit 
smaller NLO corrections (below 10\%)~\cite{pipi}.

To include $K$- and $\eta$-mesons dynamically, one must use
three-flavour $\chi$PT around the $N_f=3$ chiral limit.
Strange sea-quark loops may affect chiral series by damping
the leading-order (LO) term, which depends on 
$F^2(3)$ and $\Sigma(3)$, and by enhancing
next-to-leading-order (NLO) corrections, in
particular when violating 
the Zweig rule in the scalar sector.
Take for instance:
\begin{equation} \label{eq:fpi}
F_\pi^2 = F(3)^2 + 16(m_s+2m)B_0 \Delta L_4+ 16mB_0\Delta L_5 +O(m_q^2)
\end{equation}
where $B_0=-\lim_{m_u,m_d,m_s\to 0} \langle\bar{u}u\rangle/F_\pi^2$, and 
we have put together NLO low-energy constants and chiral logarithms
$\Delta L_5=L_5(M_\rho)+0.67\cdot 10^{-3}$,
$\Delta L_4=L_4(M_\rho)+0.51\cdot 10^{-3}$ (enhanced by $m_s$). If we assume
that the LO contribution is numerically dominant (i.e., $F_\pi^2 = F(3)^2$ to
a very good approximation), we can perform the following manipulations:
 \begin{equation} \label{eq:fpinum}
\frac{F(3)^2}{F_\pi^2}=\frac{F^2(3)}{F^2(3)+O(m_q^2)}
  = 1- 8\frac{2M_K^2+M_\pi^2}{F_\pi^2}\Delta L_4
     - 8\frac{M_\pi^2}{F_\pi^2}\Delta L_5 + O(m_q^2)
  =1- 0.51  -0.04 + O(m_q^2)
\end{equation}
where we have used $1/(1+x)=1-x$ and eq.~(\ref{eq:fpi}) at the second step,
and the second and third terms of the last equality are obtained using
$L_4(M_\rho)=0.5\cdot 10^{-3}$ and $L_5(M_\rho)=1.4\cdot 10^{-3}$ respectively~\cite{pika}. This
is clearly in contradiction with the orginal 
assumption $F_\pi^2 \simeq F(3)^2$. 
The available dispersive estimates of $L_6$~\cite{uuss} yield
a similar situation for $F_\pi^2M_\pi^2$~cite{resum}.
Therefore, a 
small positive value of $L_4(M_\rho)$ or $L_6(M_\rho)$ is enough to
spoil the rapid convergence of $N_f=3$ chiral
series and to induce a
numerical competition between formal LO and NLO contributions.

\begin{figure}[t]
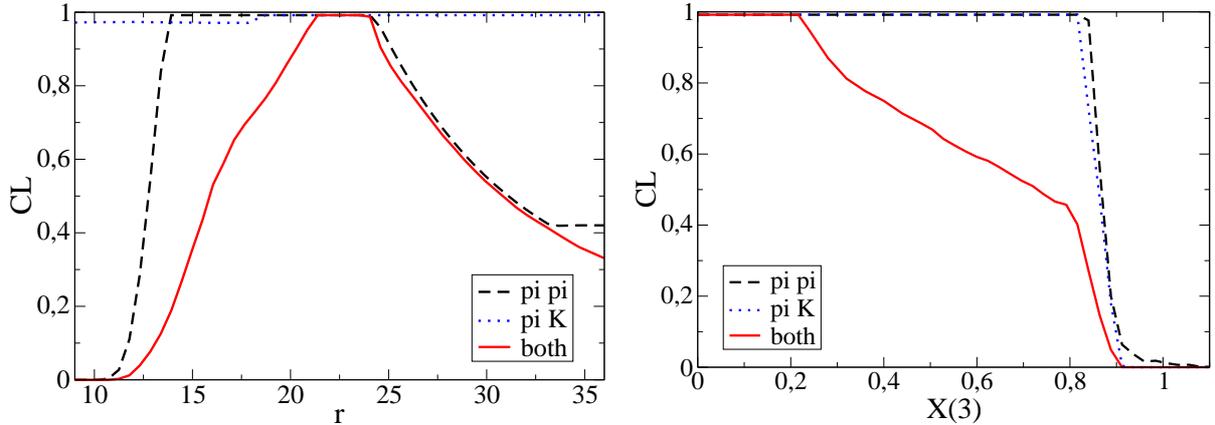

\includegraphics[height=5.6cm]{pr-1-col}
\hspace{0.1cm}
\includegraphics[height=5.6cm]{px-1-col}
\caption{Profiles for the confidence levels of $r=m_s/m$ (left) and
$X(3)=2m\Sigma(3)/(F_\pi^2M_\pi^2)$ (right). In each case, the results 
are obtained from $\pi\pi$ scattering only (dashed line), $\pi K$ scattering
only (dotted line), or both sources of information (solid line).}
\label{fig:CLs}
\end{figure}

\section{Constraints from $\pi\pi$ and $\pi K$ scatterings}

The potentially ``large'' values of 
$L_4$ and $L_6$ lead to a numerical competition 
between formal LO and NLO contributions in chiral series. To deal
with such a situation,
we have introduced a framework, called Resummed Chiral Perturbation
Theory (Re$\chi$PT)~\cite{rechpt}, where we define
the appropriate observables to consider and the treatment 
of their chiral expansion~\cite{rechpt,resum}. It allows for a
a resummation of the potentially large effect of the 
Zweig-rule violating couplings $L_4$ and $L_6$~\cite{rechpt,resum}.
Since this framework copes with the
possibility of a numerical competition between (formal) LO and NLO terms in
chiral series, some usual $O(p^4)$ results are not valid any longer: 
for instance $r=m_s/m$ is not fixed by $M_K^2/M_\pi^2$ and 
may vary from 8 to 40.

One can apply this framework 
to $\pi\pi$ and $\pi K$ scatterings, which provide information on $N_f=2$ and
$N_f=3$ patterns of chiral symmetry breaking respectively, and in particular on
$r=m_s/m$, the quark condensate
$X(3)=2m\Sigma(3)/(F_\pi^2M_\pi^2)$ and the decay constant
$Z(3)=F^2(3)/F_\pi^2$.
One can exploit dispersive relations, such
as Roy equations~\cite{royeq} and Roy-Steiner equations~\cite{pika}, to reconstruct the amplitudes from the phase shifts 
from threshold up to energies around 1 GeV.
Matching the dispersive and chiral representations of the amplitude in a
frequentist framework provides constraints (in terms of confidence levels) 
on the main parameters of interest for three-flavour $\chi$PT~\cite{rechpt}. 
Fig.~\ref{fig:CLs} shows the situation for $r=m_s/m$ and 
$X(3)=2m\Sigma(3)/(F_\pi^2M_\pi^2)$.

The main impact of $\pi\pi$ scattering consists in constraining $r=m_s/m$:
indeed, $\pi\pi$ scattering 
pins down the two-flavour condensate $X(2)$ 
rather accurately, which can be related to $r$ through
the spectrum of pseudoscalar mesons~\cite{param,resum}. 
The combination of $\pi\pi$ and $\pi K$ scatterings yields: 
\begin{equation}
r \geq 14.8\,, \quad  X(3)\leq 0.83\,, \quad Y(3)\leq 1.1\,, 
  \quad 0.18 \leq Z(3)\leq 1\,. \qquad  [68 \% {\rm CL}]
\end{equation}

\section{Conclusion}
The presence of massive $s\bar{s}$-pairs in the QCD vacuum may induce
significant differences in the pattern of chiral symmetry breaking
between the $N_f=2$ and $N_f=3$ chiral limits.
This effect, related to the violation of the Zweig rule in the scalar
sector, may spoil the convergence of three-flavour chiral expansions.
We introduce
Resummed Chiral Perturbation Theory to deal with such a problem, and apply
it to our experimental knowledge on $\pi\pi$ and $\pi K$ scatterings. The
outcome does not favour the usual picture 
of a large quark condensation independent of the number
of massless flavours. Further experimental information is needed 
to constrain the pattern of chiral symmetry 
breaking efficiently
and learn more about its variation with $N_f$.

\ack

Work supported in part by the EU Contract No. MRTN-CT-2006-035482, \lq\lq
FLAVIAnet''.

\end{document}